\documentclass[10pt]{article}
\usepackage{graphicx}

\newcommand{\be}{\begin{equation}}
\newcommand{\ee}{\end{equation}}
\newcommand{\bea}{\begin{eqnarray}}
\newcommand{\eea}{\end{eqnarray}}

\begin{document}
\begin{center}
\Large
{\bf Cosmological perturbations in Palatini modified gravity}
\vspace{.3in}
\normalsize

\large{Kotub Uddin, James E. Lidsey and Reza Tavakol}

\normalsize
\vspace{.3in}

{\em Astronomy Unit, School of Mathematical Sciences, \\
Queen Mary, University of London, London, E1 4NS, U.K.}

\vspace{.1in}

K.Uddin@qmul.ac.uk, J.E.Lidsey@qmul.ac.uk, R.Tavakol@qmul.ac.uk

\end{center}

\begin{abstract}

\noindent Two approaches to the study of cosmological density perturbations 
in modified theories of Palatini gravity have recently been discussed. 
These utilise, respectively, a generalisation of Birkhoff's theorem 
and a direct linearization of the gravitational field equations.
In this paper these approaches are compared and contrasted.   
The general form of the gravitational lagrangian 
for which the two frameworks yield identical results in the  
long-wavelength limit is derived. This class of models includes the case where 
the lagrangian is a power-law of the Ricci curvature scalar. 
The evolution of density perturbations 
in theories of the type $f(R)=R-c /R^ b$ is investigated numerically. 
It is found that the results obtained by the two methods
are in good agreement on sufficiently 
large scales when the values of the parameters $(b,c)$ are 
consistent with current observational constraints. However, this agreement 
becomes progressively poorer for models that differ significantly 
from the standard concordance model and as smaller scales are considered. 
\end{abstract}

\section{Introduction}

There is now overwhelming 
observational evidence that the universe is currently undergoing
a phase of accelerated expansion. This evidence arises from
high redshift supernovae surveys \cite{Rie98,Per99,DalDjo03,Rie04,Ast06},
observations of large-scale structure
\cite{Tegmark-etal04,Seljak-etal05}, baryon acoustic
oscillations \cite{Eis05}, and high-precision
data from the Cosmic Microwave Background (CMB) \cite{Spe03,Spe06}.

An accelerating universe poses
a great challenge to modern cosmology. It is very difficult 
to explain such behaviour within a conventional 
general relativistic framework.
The simplest way to generate a phase of 
acceleration is to introduce a
cosmological constant into the field equations. Although this is 
entirely consistent with the available data, the microphysical 
origin of such a term remains a mystery.  

In view of this, a large number of alternative
models have been proposed (see \cite{Copeland:2006wr} for a recent review). 
In most cases, these can be classified into two broad groups: 
those that invoke an exotic matter source for the dark energy \cite{Peebles:2002gy} 
and those that modify the gravitational
sector of the theory \cite{Dvali:2000hr}. Examples of the latter include
generalised theories of gravity 
based on non-linear functions, $f(R)$, of the Ricci scalar $R$. 
Such modifications to the (linear) Einstein-Hilbert action typically arise
in effective actions derived from string/M-theory
\cite{Buchbinder-et-al-1992,Gasperini-Venezinano-1992,Noriji-etal-2003,Vassilevich-2003}.

Given the large number of candidates within both approaches,
it is important to determine whether observations can 
be employed to discriminate between them. In general, 
the homogeneous dynamics encoded in the modified 
Friedmann equations of $f(R)$ theories can always be 
expressed in terms of a conventional relativistic cosmology 
sourced by an effective perfect fluid. Thus, it is necessary to include 
inhomogeneities in order to lift the degeneracy between the two 
frameworks. This can be achieved, for example, 
by considering the evolution of density perturbations and 
a study of perturbation theory in $f(R)$ gravity is therefore 
of considerable interest. Two approaches
have been developed recently in this context. One possibility 
is to follow the standard procedure 
employed in relativistic perturbation theory 
and derive the perturbation equations by linearizing the gravitational 
field equations \cite{kks06,Hwang:2001qk}.
On the other hand, an alternative procedure 
has recently been put forward by Lue, Scoccimarro and Starkman (LuSS) 
\cite{lue03} which employs a generalised version of Birkhoff's theorem
(see also Ref. \cite{mult}). 
This procedure has the benefit of greatly simplifying the analysis, but 
suffers from the drawback that the degree of its applicability in
more general settings is presently not known in detail.

The purpose of the present work is to perform a 
detailed comparative study of the evolution of perturbations obtained
by employing the LuSS procedure and the direct
linearization of the field equations. Such a comparison can 
serve as a crucial step in clarifying the status of the 
LuSS approach in non-linear gravity theories. 

When studying generalised $f(R)$ gravity,
a choice must be made as to which independent fields should 
be varied in the action. There are two possibilities, 
which are referred to as the `metric' and `Palatini' frameworks, 
respectively. In the former case, only variations  
with respect to the metric are considered, whereas the action 
is varied with respect to both the metric and the connection 
in the Palatini framework \cite{Vollick:2003aw}. Both approaches result 
in identical field equations for the 
Einstein-Hilbert action, where $f(R) \propto R$. 
More generally, 
however, the metric approach results in fourth-order equations, whereas the Palatini variation
generates a second-order system \cite{Meng:2003ry}. Moreover, the theories
of the type $f(R)=R-c/R^b$ based on metric formalism have difficulty in passing the solar system tests
\cite{Chiba:2003} (see, however, \cite{Hu07}) and producing the correct Newtonian limit \cite{Sotiriou:2006}. 
Such theories also suffer from gravitational instabilities as discussed in \cite{Dolgov:2003}. Along with these 
concerns, a recent study has found that these types of theories can not produce a standard matter epoch
\cite{Amendola:2007}. In this paper, we consider theories
based on the Palatini variational method which can avoid the problems outlined above. Such theories have 
received considerable attention  in recent years as candidates for explaining 
the present-day acceleration of the universe 
\cite{kks06, Vollick:2003aw, Meng:2003ry, Palatiniclassics, Meng04,  mota06, Capo03, Carroll03, reza07}.

The outline of the paper is as follows.
In section 2, we briefly review the Palatini variational method 
and present the perturbation evolution equations 
derived by using the LuSS procedure and the direct linearization 
of the field equations. In section 3,  we derive the 
necessary conditions on the 
general form of the gravitational lagrangian $f(R)$  
for the two approaches to be compatible.  
We perform a numerical analysis in section 4 
to quantitatively compare the 
two frameworks for the class of 
$f(R)$ theories where the action includes
an inverse power of the Ricci curvature scalar. 
We conclude with a discussion in Section 5. 

\section{$\mathbf{f(R)}$ gravity in the Palatini formalism}

\subsection{The Field Equations}
 
We consider the class of non-linear gravity
theories defined by the action
\begin{equation}
\label{actpal}
S=  \frac{1}{2\kappa}\int d^4x\sqrt{-g} \left[ f (R ) \right] +S_m (g, \psi)   ,
\end{equation}
where $\kappa \equiv 8\pi G$ is a constant, 
$f(R)$ is a differentiable function of
the Ricci scalar $R \equiv g^{\mu\nu}R_{\mu\nu} (g, \hat{\Gamma})$
and $R_{\mu\nu} (g, \hat{\Gamma})$ is the Ricci tensor of the 
affine connection $\hat{\Gamma}^{\alpha}_{\beta\gamma}$. 
The matter action $S_m$ is a functional only of the metric tensor and matter fields, $\psi$.

In the Palatini framework, the 
affine connection and the metric are 
treated as independent variables. 
Extremizing action (\ref{actpal}) with respect to the metric tensor 
yields the gravitational field equations (see, e.g., Ref.~\cite{Vollick:2003aw}): 
\be
\label{Gpal}
F(R) R_{\mu\nu}-\frac{1}{2} f(R) g_{\mu\nu}= \kappa T_{\mu\nu}, 
\ee
where $F (R) \equiv d f/d R$ and 
\begin{equation}
T_{\mu\nu} = -\frac{2}{\sqrt{-g}}
\frac{\delta S_m}{\delta g^{\mu\nu}}
\end{equation}
defines the energy-momentum tensor. 
Contraction of Eq. (\ref{Gpal}) yields the algebraic constraint 
equation
\be
\label{structural}
RF(R)- 2f(R) =\kappa T ,
\ee
where $T \equiv {T^{\mu}}_{\mu}$. 

On the other hand, varying  action (\ref{actpal}) 
with respect to the affine connection 
$\hat{\Gamma}^{\alpha}_{\beta\gamma}$ and contracting 
implies that \cite{Vollick:2003aw}
\be
\label{connection}
\hat{\nabla}_{\rho}[F(R)\sqrt{-g}g^{\mu\nu}]=0  ,
\ee
where $\hat{\nabla}$ is the covariant derivative 
defined by $\hat{\Gamma}^{\alpha}_{\beta\gamma}$. 
The solution to Eq. (\ref{connection}) 
is given by writing $\hat{\Gamma}^{\alpha}_{\beta\gamma}$ as 
the Levi-Civita connection for a new metric 
$h_{\mu \nu} \equiv F(R) g_{\mu \nu}$, 
which is conformally equivalent to the spacetime metric $g_{\mu \nu}$.
As a result, we may write the Ricci tensor of the affine connection  
in the form 
\be
\label{ricci}
R_{\mu\nu} (\hat{\Gamma}) 
=R_{\mu\nu}(g)+ \frac{3}{2} 
\frac{\nabla_{\mu} F \nabla_{\nu} F}{F^2}
-\frac{1}{F} \nabla_{\mu\nu}F
-\frac{1}{2F}g_{\mu\nu} \nabla_{\lambda} \nabla^{\lambda}F  ,
\ee
where $R_{\mu\nu}(g)$ is the Ricci tensor of the Levi-Civita connection. 
It can then be shown that the
field equations (\ref{Gpal}) and constraint equation (\ref{structural})
can be derived from a scalar-tensor action of the 
form \cite{Wang:2004pq}
\begin{equation}
\label{bdaction}
S= \frac{1}{2\kappa} \int d^4x \sqrt{-g} \left[ \phi R(g) 
+ \frac{3}{2\phi} \left( \nabla \phi \right)^2 -V(\phi ) \right] 
+S_m  ,
\end{equation}
where the scalar field
$\phi \equiv F$ and $V(\phi ) \equiv 
R(\phi )F -f[R(\phi)]$ represents the potential. Action 
(\ref{bdaction}) corresponds to the Brans-Dicke theory 
with a dilaton-graviton 
coupling $\omega_0 =-3/2$. 

We assume throughout that the background spacetime is 
given by the spatially flat and isotropic 
Friedmann-Robertson-Walker (FRW) metric  
\be
\label{conformal}
ds^2=a(\tau)^2(-d\tau^2+\delta_{ij}dx^i dx^j) ,
\ee 
where $a(\tau )$ is the scale factor of the universe 
and $\tau \equiv \int dt /a(t)$ defines conformal time. 
We further assume that the matter content of the universe is 
represented by a pressureless perfect fluid with energy-momentum tensor
${T^{\mu}}_{\nu} = {\rm diag} (-\rho_m , 0, 0, 0)$. 
The  Ricci tensor (\ref{ricci}) may then be employed to derive 
the generalised Friedmann equation for this cosmology: 
\be
\label{Freidmann1}
6F\left(H+\frac{\dot{F}}{2F}\right)^2 - f=\kappa \rho_{m}  ,
\ee
where a dot denotes differentiation with respect to coordinate time $t$
and $H\equiv \dot{a}/a$ defines the Hubble expansion parameter.
After substituting the trace equation (\ref{structural}), 
the generalised Friedmann equation can be rewritten in the form 
\be
\label{H2}
H^2=\frac{3f-RF}{6F}
\left( 1-\frac{3}{2}\frac{F_{,R}(2f-RF)}{F(F-RF_{,R})}\right)^{-2}  ,
\ee
where a comma denotes differentiation. 

\subsection{The Perturbation Evolution Equations}

In conventional cosmology, 
there exists an interesting equivalence between the 
Newtonian and general relativistic frameworks. Both approaches 
result in identical background
evolution equations (i.e. Friedmann equations) as well as evolution 
equations for the scalar perturbations.  
The former coincidence results from the fact that there is an analogue of  
Newton's sphere theorem in general relativistic settings, 
i.e., Birkhoff's theorem holds. The correspondence for the 
perturbation evolution arises in the absence of 
vector and tensor fluctuations.

Recently, a procedure has been put forward
by Lue, Scoccimarro and Starkman \cite{lue03}
which relies on the assumption that this Newtonian analogy,
including Birkhoff's theorem,
holds in the more general setting of modified gravity theories.
According to this procedure, it is assumed that 
the growth of large-scale structure can be modelled 
in terms of a uniform sphere of dust of constant
mass, such that the evolution inside the sphere is determined by
the FRW metric. Using Birkhoff's theorem, 
the spacetime metric in the empty exterior is then taken
to be Schwarzschild-like. 
The components of the exterior metric 
are then uniquely determined by smoothly matching  
the interior and exterior regions. 

The overdensity $\delta (t)$ 
of the spherical distribution of pressureless matter with mass $M$ 
and radius $r$ is defined by 
\begin{equation}
1+ \delta (t) \equiv \frac{3M}{4\pi \rho r^3}  ,
\end{equation}
where $\rho (t)$ represents the background energy density. The matching
conditions imply that $\ddot{r}=r(H^2+\dot{H})$ and the evolution 
of the density perturbation is then given by \cite{lue03, luefake}
\be
\label{lueode1}
\ddot{\delta}+2H\dot{\delta}-\left(2\dot{H}+\frac{\ddot{H}}{H}\right)\delta=0  ,
\ee
or, equivalently, by  
\be
\label{birkhoffequation}
\delta''+\mathcal{H}\delta'-\left(\frac{\mathcal{H}''}{\mathcal{H}}-2\mathcal{H}'\right)
\delta=0 ,
\ee
where a prime denotes differentiation with respect to conformal time
and $\mathcal{H} \equiv aH =\dot{a}$. Eq. (\ref{birkhoffequation}) 
can also be derived by assuming that the continuity and Friedmann equations 
apply directly to the fluctuations \cite{mult}. 

Recently, the evolution of perturbations in $f(R)$ gravity 
was investigated using the LuSS procedure \cite{lue03,mota06}. 
The advantage of this approach is that 
the growth of the density contrast can be expressed in terms of a single 
quadrature involving the Hubble parameter and the scale factor  
\cite{lue03,heath}: 
\begin{equation}
\label{quadlue}
\delta \propto H \int \frac{dt}{a^2H^2}  .
\end{equation}
In principle, therefore, the evolution 
of the perturbations can be determined 
once the background dynamics has been specified. 

However, the validity of the LuSS procedure has yet to be established 
in generalised gravity. It is important, therefore, to 
compare this approach with the method that directly linearises 
the gravitational field equations. The corresponding evolution 
equation for comoving, linear density perturbations in a pressureless
universe was recently derived by Koivisto and Kurki-Suonio (KKS)
using this direct method and 
found to have the form \cite{kks06} 
\bea
\label{kksequation}
\delta''&+&
3\mathcal{H}\frac{2F\mathcal{H}(F\mathcal{H}^2+F'')-2F'^{2}\mathcal{H}+F'F(-2\mathcal{H}'+\mathcal{H}^2)}{3F\mathcal{H}^2(2F\mathcal{H}+F')}\delta' \\
&-&\frac{6F^2\mathcal{H}^2(\mathcal{H}''-2\mathcal{H}'\mathcal{H})+6F'^2\mathcal{H}(\mathcal{H}^2-\mathcal{H}')+F'F(3\mathcal{H}''\mathcal{H}-6\mathcal{H}'^2-\mathcal{H}^2k^2)+6F''F\mathcal{H}(\mathcal{H}'-\mathcal{H}^2)}{3F\mathcal{H}^2(2F\mathcal{H}+F')}\delta=0 ,
\nonumber
\eea
where $k$ is the comoving wavenumber that arises due to the 
Fourier decomposition. 

We will refer to Eqs. (\ref{birkhoffequation}) and (\ref{kksequation}) 
as the LuSS and KKS perturbation
equations, respectively. 
We will be interested in identifying the 
domain where the equation based on the 
LuSS procedure provides an accurate description for the 
evolution of the perturbations. In the following section, we 
adopt an analytical approach with the aim of 
identifying the general form of the gravitational lagrangian, 
$f(R)$, for this to be the case.

\section{Analytical Results} 

A direct comparison between the LuSS equation (\ref{birkhoffequation}) 
and the KKS equation (\ref{kksequation}) suggests that the latter  
should be rewritten in the form 
\be
\label{kksode}
\delta''+\xi \mathcal{H}\delta'-
\zeta \left(\frac{\mathcal{H}''}{\mathcal{H}}-2\mathcal{H}'\right)\delta=0,
\ee
where the parameters $\xi$ and $\zeta$ are defined by
\be
\label{xi}
\xi \equiv 1+\frac{2FF''\mathcal{H}-2F'^2\mathcal{H}-2FF'\mathcal{H}'}{F\mathcal{H}^2(2F\mathcal{H}+F')},
\ee
and 
\be
\label{newzeta}
\zeta \equiv 1+
\frac{\mathcal{H}^2-\mathcal{H}'}{\mathcal{H}''-2\mathcal{H}'\mathcal{H}}(1-\xi)-
\frac{F'\mathcal{H}}{3(2F\mathcal{H}+F')(\mathcal{H}''-2\mathcal{H}'\mathcal{H})}k^2 ,
\ee
respectively. The form of Eq. (\ref{kksode}) implies that
the LuSS and KKS equations are equivalent 
when $\xi = \zeta =1$, but it is clear that this occurs 
only for Einstein gravity where $F'=0$. 
Indeed, the most striking 
difference is the presence of the gradient term in the KKS equation. 
Such a term also arises in the corresponding density perturbation
equation derived in the metric variational approach \cite{k-terms met}.
The origin of this term can be understood from the dynamical equivalence 
between Palatini gravity and Brans-Dicke theory, as expressed in 
Eq. (\ref{bdaction}). Fluctuations in the pressureless matter 
induce perturbations in the scalar field $\phi$ (i.e. the Ricci curvature), 
which in turn generate a pressure gradient in the fluid. In general,
the sound speed of the fluctuations in the cold dark matter is given by 
\begin{equation}   
\label{soundspeed}
c_s^2 = \frac{F'}{3(2F\mathcal{H} +F')}  .
\end{equation}

The magnitude of 
$\xi$ is independent of $k$ and is therefore unaffected by the specific choice 
of scale. However, $\zeta$ contains a 
gradient term which is  proportional to $k^2$ and this may be 
significant on small scales. Consequently,   
the evolution of the perturbations will indeed be different
in the two approaches. However, the gradient term becomes 
negligible in the long-wavelength limit (which corresponds formally to 
$k^2 \rightarrow 0$). In this limit, a necessary and sufficient 
condition for equivalence between the LuSS and KKS equations 
is that $\xi =1$ and this constraint is satisfied when  
\begin{equation}
\label{neccon}
FF''\mathcal{H}-F'^2\mathcal{H}-FF'\mathcal{H}' =0  .
\end{equation}

Eq. (\ref{neccon}) may be viewed as a second-order, non-linear 
differential equation for $F (\tau )$. One solution to this equation 
is that of general relativity with a cosmological constant, 
$f(R)=R-\Lambda$. More generally, if $F' \ne 0$ and $F'' \ne 0$, 
we may define a 
parameter $Y \equiv F'/F$. 
This reduces Eq. (\ref{neccon}) to the remarkably 
simple form
\begin{equation}
\label{remarkablysimple}
\frac{Y'}{Y} = \frac{\mathcal{H}'}{\mathcal{H}}  ,
\end{equation}
which admits the integral $Y=Y_0 \mathcal{H}$, where $Y_0$ is an arbitrary 
integration constant. This in turn implies that 
\begin{equation}
\label{Fsolution}
F=F_0 a^{Y_0}  ,
\end{equation}
where $F_0$ is a second integration constant. 

On the other hand, the trace equation (\ref{structural}) for a universe 
sourced by pressureless matter reduces to the condition \cite{mota06}
\begin{equation}
\label{af}
a \propto \left( 2f -R \frac{df}{dR} \right)^{-1/3}  .
\end{equation}
Hence, substitution of Eq. (\ref{af}) into Eq. (\ref{Fsolution}) 
yields a first-order, non-linear differential equation in 
the gravitational lagrangian $f(R)$: 
\begin{equation}
\label{lagrange}
\left( \frac{df}{dR} \right)^n  \left( 2f -  R \frac{df}{dR} \right) = 
{\rm constant}  ,
\end{equation}
where $n \equiv 3/Y_0$. 

Eq. (\ref{lagrange}) is a  particular example of 
d'Alembert's equation and may be solved in full generality 
\cite{zwillinger}.  
Since we are interested in the functional dependence of 
the lagrangian on the Ricci scalar, we may rescale $f$ without 
loss of  generality 
such that the constant on the right-hand side of Eq. (\ref{lagrange}) 
is unity. If we now define the functions
\begin{equation}
\label{defMN}
M \equiv \frac{1}{2} \frac{df}{dR} , \qquad 
N \equiv \frac{1}{2} \left( \frac{df}{dR} \right)^{-n}
\end{equation}
and denote $p \equiv df/dR$,  
Eq. (\ref{lagrange}) can be expressed in the form 
$f(R) = R M(p ) + N(p)$. Differentiating this expression 
with respect to $R$ then yields  
\begin{equation}
\label{towardslinear}
p = M(p) + \frac{dp}{dR} \left[ R 
\frac{dM(p)}{dp} +\frac{dN(p)}{dp} \right]  .
\end{equation}

However, Eq. (\ref{towardslinear}) can be expressed as 
a linear differential equation in the dependent variable $R$ 
and independent variable $p$: 
\begin{equation}
\label{linear}
\frac{dR}{dp} - \frac{R}{p} = -\frac{n}{p^{2+n}}  .
\end{equation}
Hence,  solving Eq. (\ref{linear}) by the method of integrating factors 
yields the general solution to Eq. (\ref{lagrange}) in a parametric 
form: 
\begin{eqnarray}
\label{gensol1}
R= C_0 P + \frac{n}{n+2} \frac{1}{P^{1+n}}
\\
\label{gensol2}
f= \frac{1}{2} RP + \frac{1}{2P^n}  ,
\end{eqnarray}
where $C_0$ is an arbitrary integration constant and $P$ is a free 
parameter.    

Eqs. (\ref{gensol1})-(\ref{gensol2}) 
represent the general form of the gravitational 
lagrangian $f(R)$ for the LuSS and KKS equations to be compatible
in the long wavelength limit.
It is interesting that for this class of theories 
the sound speed of the fluctuations is constant with a numerical value 
given by  
\begin{equation}
\label{constantspeed}
c_s^2= \frac{1}{3+2n}  .
\end{equation}
 
When $C_0 =0$, which is equivalent to the asymptotic limit where 
$R$ is sufficiently small, the gravitational action depends 
on a simple power of the Ricci scalar: 
\begin{equation}
\label{asumptote}
f(R) \propto R^{n/(1+n)}  .
\end{equation}
For this class of theories the Friedmann equation (\ref{H2}) 
reduces to 
\be
\label{Hpowerlaw}
H^2=\frac{3+2n}{6n}\left(1+\frac{3}{2n} \right)^{-2} R  ,
\ee
which in turn implies that the background dynamics 
is given by a power-law solution for the scale factor, 
$a \propto \mathcal{H}^{-2n/(3+n)} \propto \eta^{2n/(3+n)}$
\cite{powerlaw}. Consequently, the cosmic dynamics is equivalent to 
that of a conventional relativistic universe dominated by a perfect 
fluid with a constant equation of state. 
Finally, the parameter $\zeta$ simplifies in this case to 
\be
\label{xzpowerlaw}
\zeta=1-\frac{2n^2}{3(1+n)(3+n)(3+2n)}\frac{k^2}{\mathcal{H}^2}.
\ee

In conclusion, therefore, the above analysis indicates 
that the LuSS equation 
should provide a good approximation to the full evolution
equation for the linear density perturbation on sufficiently 
large scales in any modified gravity theory 
that asymptotes in the low-energy limit 
to a power-law in the Ricci curvature scalar. On the other hand, 
for fixed values of $n$ and $\mathcal{H}$, the LuSS equation 
becomes progressively less accurate as we move to 
smaller scales (i.e. as $k$ increases). In the following section, we 
will quantify these conclusions further by performing numerical 
calculations for a specific class of modified gravity   theories. 

\section{Numerical Results}

Motivated by the results of the previous Section, we consider  
the class of gravity theories defined by 
\be
\label{theories}
f(R) = R-\frac{c}{R^b}  ,
\ee
where $b$ and $c$ are free parameters whose values are 
constrained by observations. Such theories  
have recently been considered as possible
candidates for explaining the late-time acceleration of the universe 
\cite{Capo03, Carroll03, reza07}. 
In particular, a recent study found that 
data obtained from CMB, baryon oscillation and large-scale structure 
observations constrains the parameters $(b,c)$ to lie in the ranges 
$b \in [-0.2,1.2]$ and $c \in [-3.5,6.6]$ 
at the 68\% confidence level \cite{reza07}.
The best-fit model corresponds to the values $(b,c)=(0.027,4.63)$ and 
the $\Lambda$CDM concordance model is represented by 
$(b,c)= (0,4.38)$. These values are consistent with 
the results of other studies 
that employ CMB and supernovae data \cite{mota06}.

For the above choice of parameters, 
we have made a detailed comparative study of the
evolution of the density perturbations 
for both the LuSS equation (\ref{birkhoffequation}) and the KKS equation 
(\ref{kksequation}). The results of such a comparison 
can be quantified by defining a `fractional difference' 
parameter  
\be
\label{defDelta} 
\Delta \equiv  \frac{\delta_{LuSS} -\delta_{KKS}}{\delta_{KKS}}  ,
\ee
where subscripts `LuSS' and `KKS' refer to the results 
obtained using the LuSS and KKS  equations, respectively. 
Thus, the two approaches are completely compatible when 
$\Delta =0$. This parameter is defined in such a way that 
the difference between the two approaches is 
of the same order as the KKS approach  
when $\Delta \simeq \mathcal{O} (1)$. To a first 
approximation, therefore, it is reasonable to 
suppose that the LuSS equation becomes unreliable 
when $\Delta \approx 1$. 

There are three physical parameters in the field equations 
whose values need to be specified in the numerical integrations. These are 
$\Omega_{m0}$, $R_{0}$ and $H_{0}$, where a subscript zero indicates
present-day values and $\Omega_{m}$ is the normalised matter energy 
density\footnote{Note that in modified gravity theories 
of the type considered here, this parameter need not necessarily 
be unity in a spatially flat universe.}. However, only 
two constraint equations are available, corresponding to the 
Friedmann equation (\ref{H2}) and the
trace equation (\ref{structural}). In order to be consistent, therefore,
we specify the value of $H_{0}$ to be unity, as is the usual practice
(see, e.g., \cite{mota06}).
We then use the constraint equations 
(\ref{structural}) and (\ref{H2}) 
to determine $\Omega_{m0}$ and $R_{0}$. 
The choice of Eq. (\ref{defDelta}) implies that the 
initial value of the perturbation $\delta$ is unimportant. 
Finally, we need to specify the scale of the perturbations. 
By fixing the wavenumber at a particular value, 
one focuses on perturbations 
that entered the horizon at a particular epoch. For 
illustrative purposes we consider the values $k=5$ and $k=20$,
corresponding to scales which remain within the horizon throughout our
numerical evolution.

\begin{figure}[!h]
\includegraphics[width=8cm, height=5.5cm]{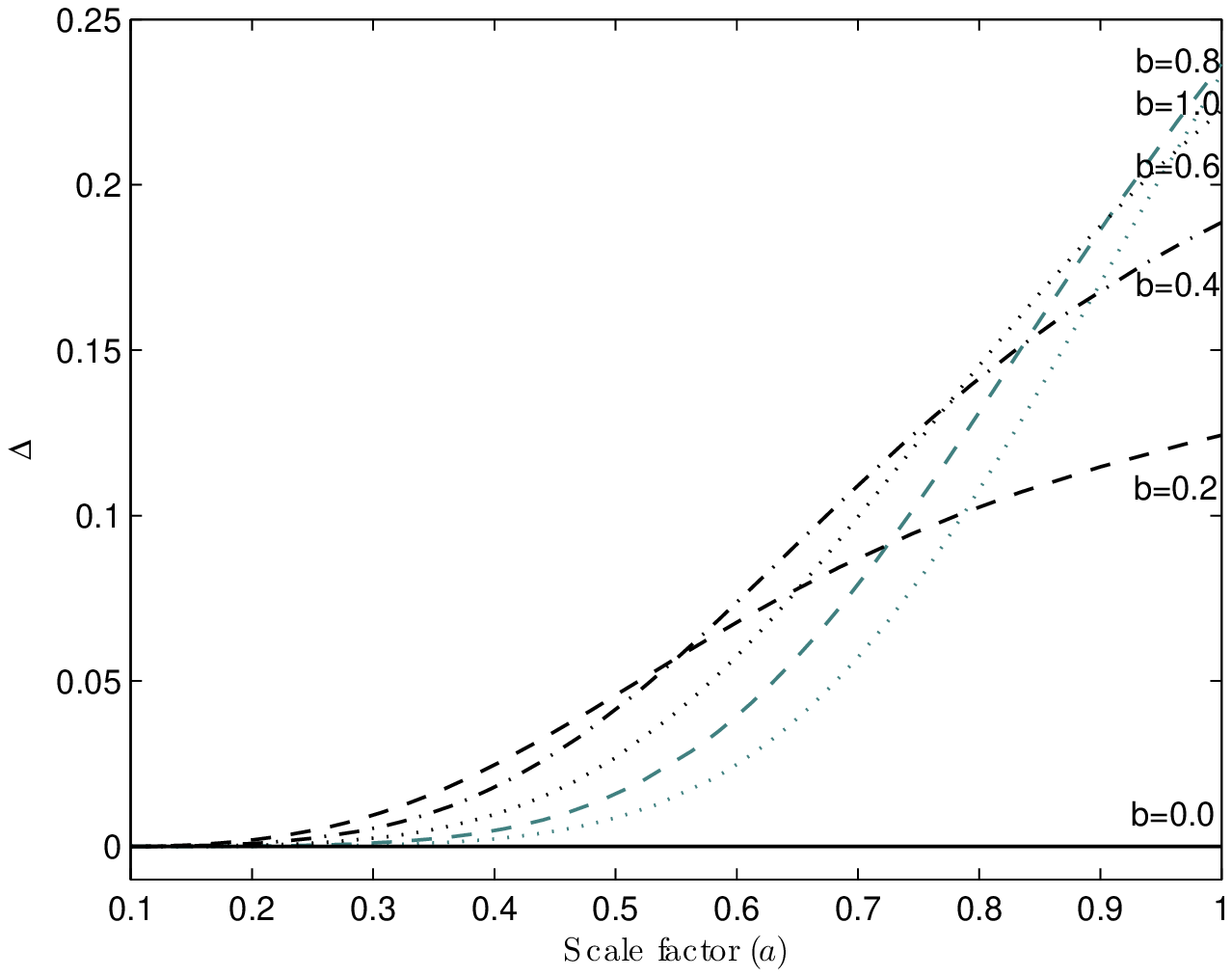}\includegraphics[width=8cm, height=5.5cm]{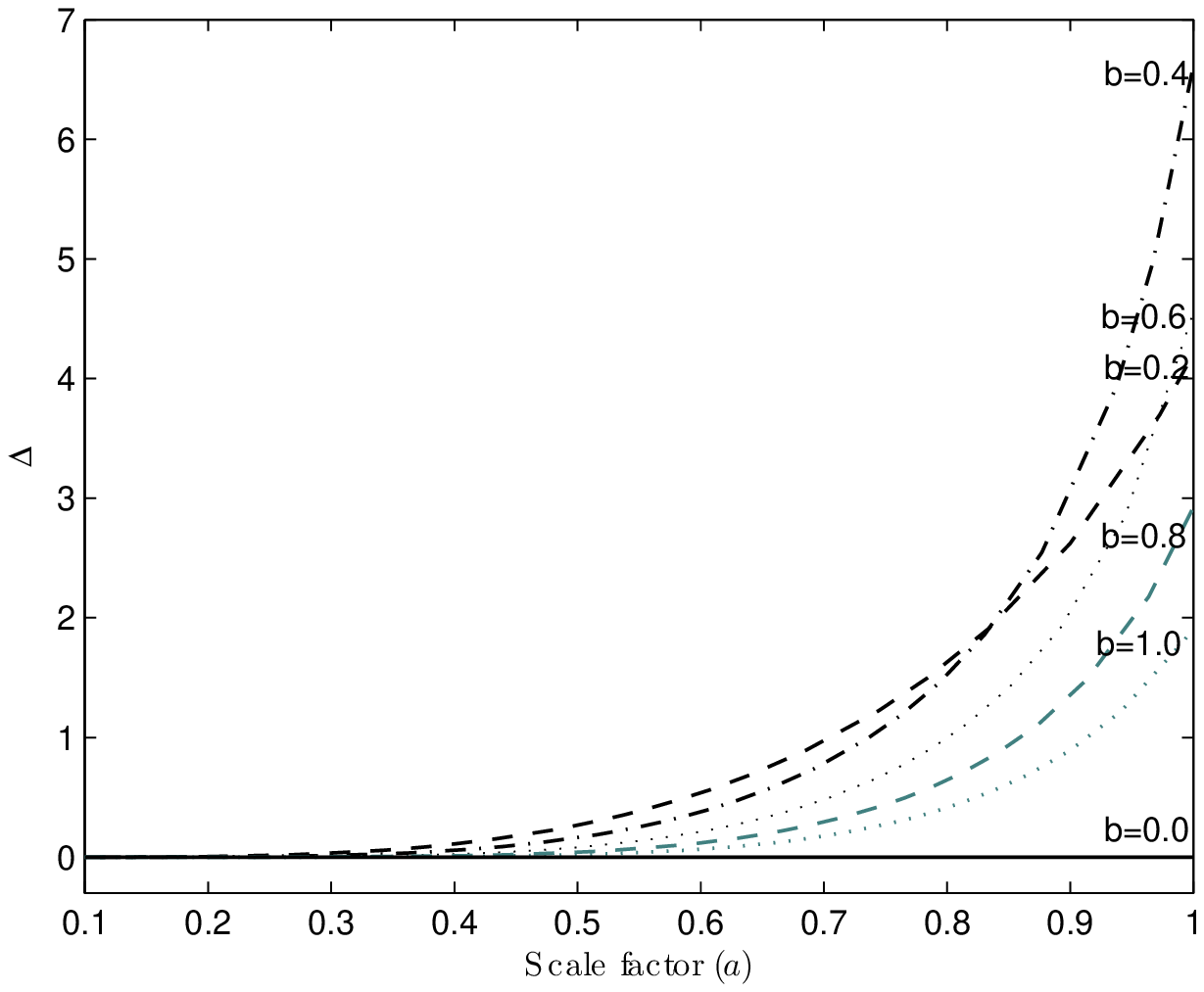}
\caption{\label{Fig1}
Illustrating how the fractional difference 
parameter, $\Delta$, varies with scale factor $a$ as
$b$ is increased. Here $c= 4.38$ and values of $b$ are 
assigned to each curve. The case $b=0$ corresponds to the $\Lambda$CDM
model. The left hand panel corresponds to $k=5$ whereas the right
hand panel corresponds to $k=20$. 
}
\end{figure}

The left hand panel of Fig. \ref{Fig1} illustrates the evolution of $\Delta$ when 
$c=4.38$ and $k=5$, with $b$ taking values in the range 
$b \in [0,1]$. As expected, $\Delta =0$ for the $\Lambda$CDM concordance 
model (given by $b=0$), 
since it is known that the LuSS equation is exact 
in this case. On the other hand, increasing the 
value of $b$ causes the behaviour of the two approaches
to deviate and the quantitative difference 
becomes more pronounced as $b$ is increased.  

We have verified that our results remain qualitatively similar  
when the parameter values lie in the ranges 
$b \in [-0.2,1.2]$ and $c \in [-3.5,6.6]$, respectively.
An important outcome of these results is that 
for values of the parameters 
consistent with recent observations, 
the agreement between the LuSS and KKS approaches is good in the 
sense that $\Delta <0.1$ for $b <0.2$. This implies that
the LuSS equation provides a good approximation to the full 
(linear) perturbation theory (for this value of $k$). 
This can be understood by noting 
that observations constrain theoretical models 
to lie close to the $\Lambda$CDM point, where it is known that 
the LuSS equation is exact. 

Further inspection of the left hand panel of Fig. \ref{Fig1}
indicates that as the value of $b$ is increased, 
the models take longer to move away from the 
$\Lambda$CDM point $\Delta = 0$, but those with 
smaller values of $b$ subsequently find it easier to approach
$\Delta = 0$ at later times. We may gain further insight into the origin 
of this behaviour by investigating the evolution of the
quantity $Q \equiv 1-F$. This vanishes at all times for 
Einstein gravity but is 
given by $Q= -bcR^{-(1+b)}$ for the class of models (\ref{theories}). 
This parameter therefore provides a measure of the deviation away 
from general relativity. 
Our numerical calculations indicate that initially
$R \approx \mathcal{O} (10^3)$ and, consequently for 
larger values of $b$, the scale factor must grow to a larger value before 
the Ricci scalar has fallen sufficiently for the correction 
term $Q$ to become dynamically significant. In other words, the onset of 
acceleration occurs at later times for larger $b$. On the other 
hand, the correction term in $f(R)$ that is proportional 
to $R^{-b}$ will become more important 
as the universe expands. The analysis of Section 3 then indicates that 
the accuracy of the LuSS equation will improve as $f(R)$ asymptotes to a
power-law form. Consequently, $\Delta$ will begin to decrease back to zero
at later times. 

We find qualitatively similar behaviour at larger values of $k$. 
The right hand panel of Fig. \ref{Fig1} illustrates the corresponding evolution of $\Delta$ when $k=20$.
As expected, models with lower values of $b$ move away from the 
$\Delta =0$ point at smaller values of the 
scale factor. The model with the lowest non-zero value of 
$b=0.2$ crosses the solutions for $b=0.4$ and $b=0.6$.  
This can be understood from Eq. (\ref{xzpowerlaw}), which
implies that the magnitude of $\zeta$ depends on the ratio
$k^2/\mathcal{H}^2 = k^2/\dot{a}^2$. 
At a formal level, therefore, increasing the value of $k$ 
is equivalent to ending the numerical calculation at a fixed 
$k$ but with a smaller value for the scale factor.  

\begin{figure}[!h]
\includegraphics[width=8cm, height=5.5cm]{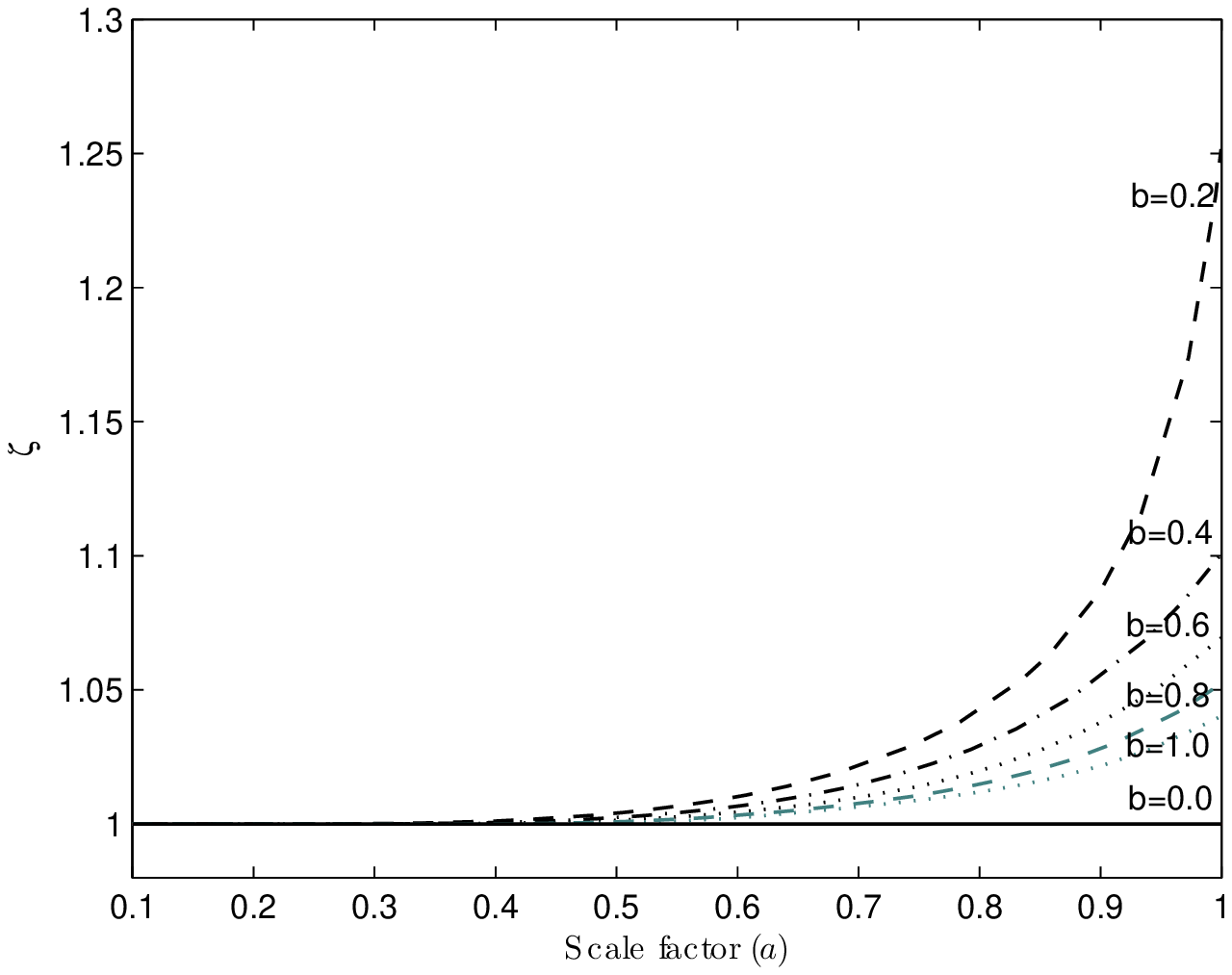}\includegraphics[width=8cm, height=5.5cm]{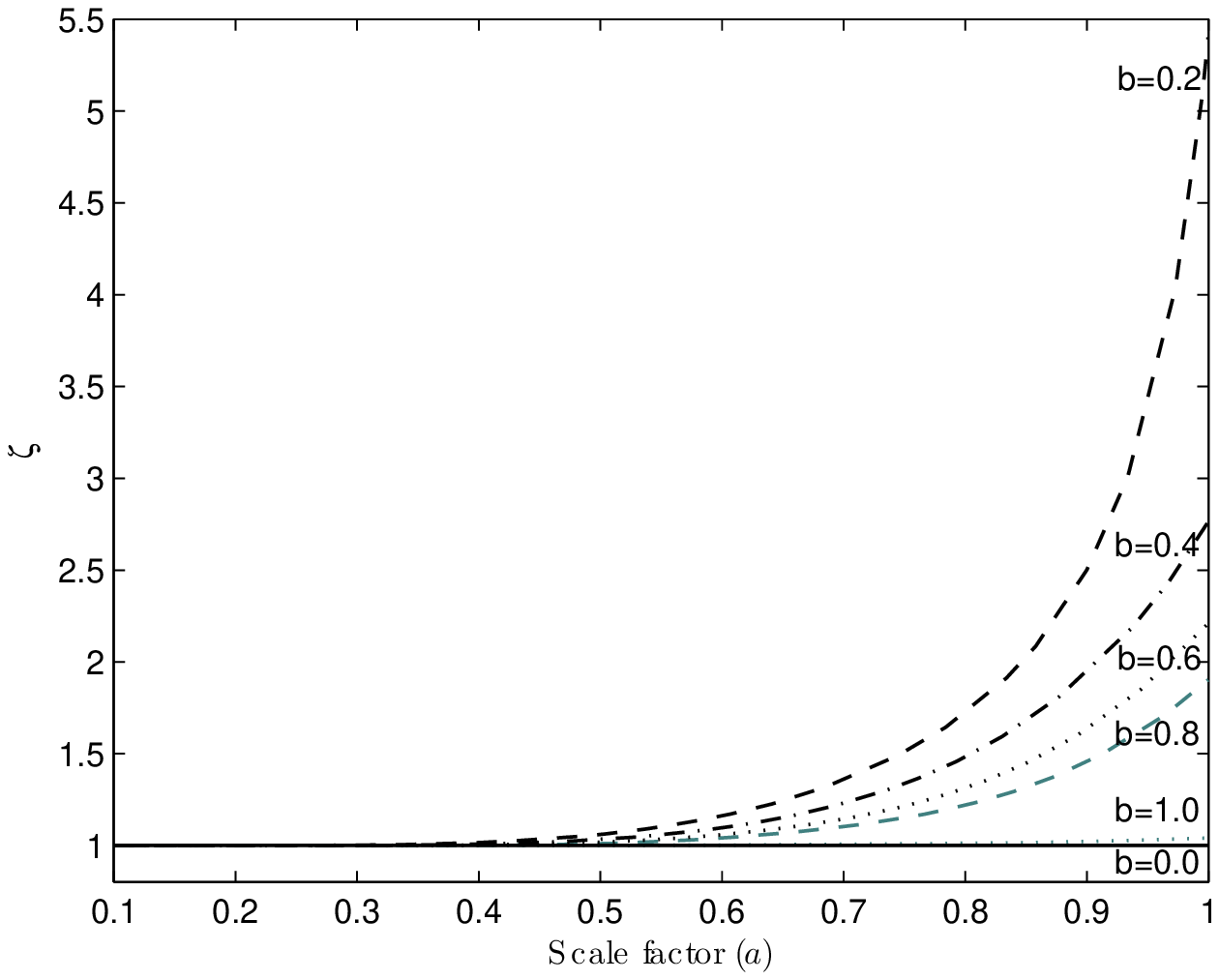}
\caption{\label{Fig2}
Illustrating the evolution of the parameter
$\zeta$ defined by Eq. (\ref{newzeta}) in the text 
for the parameter values $k=5$ (left panel) and $k=20$ (right panel). 
The LuSS procedure for
the evolution of the perturbations becomes progressively less accurate
as the deviation of this quantity from unity becomes more pronounced.}
\end{figure}

However, the quantitative agreement between the solutions 
of the LuSS and KKS equations is poor when $k=20$ and 
$\Delta$ rapidly exceeds unity in this case. This discrepancy arises 
primarily because the deviation of the parameter 
$\zeta$ away from unity is more pronounced at larger $k$. 
Fig. \ref{Fig2} illustrates the evolution of $\zeta$ for 
the different values of $k$.

\section{Conclusion}

In this paper, we have studied the evolution of density perturbations 
in generalised theories of gravity where the field equations 
are derived via the Palatini variational approach. We focused on models 
where the energy-momentum tensor is sourced by a pressureless perfect 
fluid. Two approaches to the study of density 
perturbations have recently been developed in the literature \cite{kks06, lue03, mult}. 
These involve, respectively, an application of Birkoff's 
theorem to modified gravity (the LuSS method) 
and the linearization of the full field 
equations (the KKS approach). 
In the former case, the evolution of the perturbations is determined 
entirely by the background dynamics and no pressure gradients 
are present in the perturbation evolution equation. 
However, such terms do arise in the linearization approach, which 
takes into account the fact that perturbations in the fluid induce 
fluctuations in the Ricci curvature which in turn 
modify the sound speed of the fluctuations in the matter. 
  
In the long-wavelength limit, these gradient terms are 
negligible. We have identified the most general $f(R)$ theory of gravity, 
as summarised in Eqs. (\ref{gensol1}) and (\ref{gensol2}), 
for the LuSS and KKS approaches to be compatible in this limit. 
A particular case of this class of theories arises when $f(R)$ is a
simple power law of the Ricci curvature scalar. This is interesting because
such terms are expected to arise generically as corrections to the 
Einstein-Hilbert action at low energies. 
Furthermore, theories of this type result in a background
scaling solution, in the sense 
that the homogeneous dynamics is equivalent to that of  
a conventional relativistic cosmology where the  
pressure and energy density of the perfect fluid redshift at the same rate. 
It would be interesting to explore whether this scaling behaviour 
is a necessary 
condition for compatibility between the LuSS and linearization methods
in more general theories of modified gravity. For example, a power-law 
cosmology arises in the Palatini variation of Ricci squared gravity, 
where $f \propto (R^{\mu\nu}R_{\mu\nu})^{n/2}$ \cite{Riccisquare}.

We numerically investigated a specific class of power-law
theories of the type (\ref{theories}) and compared  
the LuSS and KKS approaches  
on smaller scales where gradient terms become  
significant. We found that when 
the parameters of the underlying theory take values that are consistent 
with cosmological observations, the 
LuSS procedure provides a reasonably 
good approximation to the complete linearised theory if $k$ is 
not too large (i.e. of the order of a few or less).  
However, the agreement between the two approaches soon breaks down  
on smaller scales. 

\vspace{.1in}

\noindent {\bf Acknowledgements}

\vspace{.1in}

\noindent 
KU is supported by the Science and Technology Facilities Council (STFC). 
We thank S. Fay, C. Hidalgo and K. Malik for helpful discussions.


\end {document}